\documentclass[aps,pra,twocolumn]{revtex4}
\usepackage[latin9]{inputenc}
\usepackage{graphicx,epsf}
\usepackage{amssymb}
\usepackage{amsmath}
\usepackage{color, soul}
\usepackage{bm}
\usepackage[caption=false]{subfig}
\usepackage{placeins}

\newcommand{\w}{\omega}

\newcommand{\TN}{T_{\rm N}}

\newcommand{\mtot}{m_{\rm tot}}

\newcommand{\llangle}{\langle\!\langle}
\newcommand{\rrangle}{\rangle\!\rangle}

\newcommand{\Ztwo}{\mathbb{Z}_2}
\newcommand{\Uone}{\mathrm{U(1)}}
\newcommand{\SUtwo}{\mathrm{SU(2)}}

\newcommand{\scbo}{SrCu$_2$(BO$_3$)$_2$}

\newcommand{\absval}[1]{\lvert #1 \rvert}
\newcommand{\expval}[1]{\langle #1 \rangle}
\newcommand{\numop}[1]{#1^{\dagger} #1}

\RequirePackage[
   hyperindex,colorlinks,bookmarksnumbered,
   plainpages=true, pdfstartview=FitH,bookmarks=false
]{hyperref}
\hypersetup{linkcolor=blue,urlcolor=blue,citecolor=blue}

\definecolor{darkgreen}{rgb}{0,0.5,0}
\definecolor{darkblue}{rgb}{0,0,0.5}
\definecolor{purple}{rgb}{0.35,0,0.35}
\definecolor{orange}{rgb}{0.9,0.4,0}

\graphicspath{{./}{./figs/}}

\begin{document}
\title{
Fluctuation-induced ferrimagnetism in sublattice-imbalanced antiferromagnets\\
with application to SrCu$_2$(BO$_3$)$_2$ under pressure
}

\author{Pedro M. C\^onsoli}
\author{Max Fornoville}
\author{Matthias Vojta}
\affiliation{Institut f\"ur Theoretische Physik and W\"urzburg-Dresden Cluster of Excellence ct.qmat, Technische Universit\"at Dresden,
01062 Dresden, Germany}

\date{\today}

\begin{abstract}
We show that a collinear Heisenberg antiferromagnet, whose sublattice symmetry is broken at the Hamiltonian level, becomes a fluctuation-induced ferrimagnet at any finite temperature $T$ below the N\'eel temperature $T_{\rm N}$.
We demonstrate this using a layered variant of a square-lattice $J_1$-$J_2$ model. Linear spin-wave theory is used to determine the low-temperature behavior of the uniform magnetization, and non-linear corrections are argued to yield a temperature-induced qualitative change of the magnon spectrum.
We then consider a layered Shastry-Sutherland model, describing a frustrated arrangement of orthogonal dimers. This model displays an antiferromagnetic phase for large intra-dimer couplings. A lattice distortion which breaks the glide symmetry between the two types of dimers corresponds to broken sublattice symmetry and hence gives rise to ferrimagnetism. Given indications that such a distortion is present in the material SrCu$_2$(BO$_3$)$_2$ under hydrostatic pressure, we suggest the existence of a fluctuation-induced ferrimagnetic phase in pressurized SrCu$_2$(BO$_3$)$_2$. We predict a non-monotonic behavior of the uniform magnetization as function of temperature.
\end{abstract}

\maketitle


\section{Introduction}

The field of quantum magnetism harbors a wealth of fascinating phenomena which are driven by fluctuations \cite{lacroix_book}. These include quantum spin liquids \cite{savary_rop17,kanoda_rmp17} -- stable states of matter devoid of symmetry-breaking order --, several types of unconventional quantum phase transitions \cite{mv_rop18}, as well as a variety of symmetry-breaking states stabilized by fluctuations. A large class of the latter are described as ``order by disorder'', a mechanism where a subset of states is selected from a classically degenerate manifold by either quantum or thermal fluctuations \cite{villain80}. Order by disorder is prominent in strongly frustrated magnets, one important example being the easy-plane pyrochlore antiferromagnet where an ordered state is chosen from a one-parameter degenerate manifold \cite{zhito12}.

Among the various frustrated spin systems, the Shastry-Sutherland model \cite{shastry81} plays a prominent role. It describes a planar Heisenberg model of coupled pairs of spins $1/2$ with a particular orthogonal-dimer structure.
Its ground-state phase diagram features a dimer-singlet state, a symmetry-breaking plaquette-singlet state, and a N\'eel antiferromagnet as function of increasing ratio of inter-dimer to intra-dimer couplings, $x=J'/J$ \cite{miya99,koga00,corboz13}. Very recently, a narrow quantum spin-liquid phase has been proposed in addition \cite{yang21}.
Local moments arranged on the Shastry-Sutherland lattice appear in a number of compounds, the most prominent one being the spin-$1/2$ Mott insulator {\scbo} \cite{kag99,miya99}.
Remarkably, hydrostatic pressure can be used to tune $x$ in {\scbo}, and signatures of magnetic transitions have been detected around $1.8$\,GPa \cite{waki07,hara12,zayed14,zayed17,sakurai18,guo20,jimenez21} and $4.5$\,GPa \cite{hara12}, with various experimental aspects being under active debate \cite{guo20,zayed15}. NMR experiments \cite{waki07} yield evidence for two distinct Cu sites in the intermediate phase, suggesting two types of inequivalent dimers. Antiferromagnetic (AF) order has been detected by neutron diffraction in the high-pressure phase \cite{hara14}.

\begin{figure}[b]
\centerline{\includegraphics[width=\columnwidth]{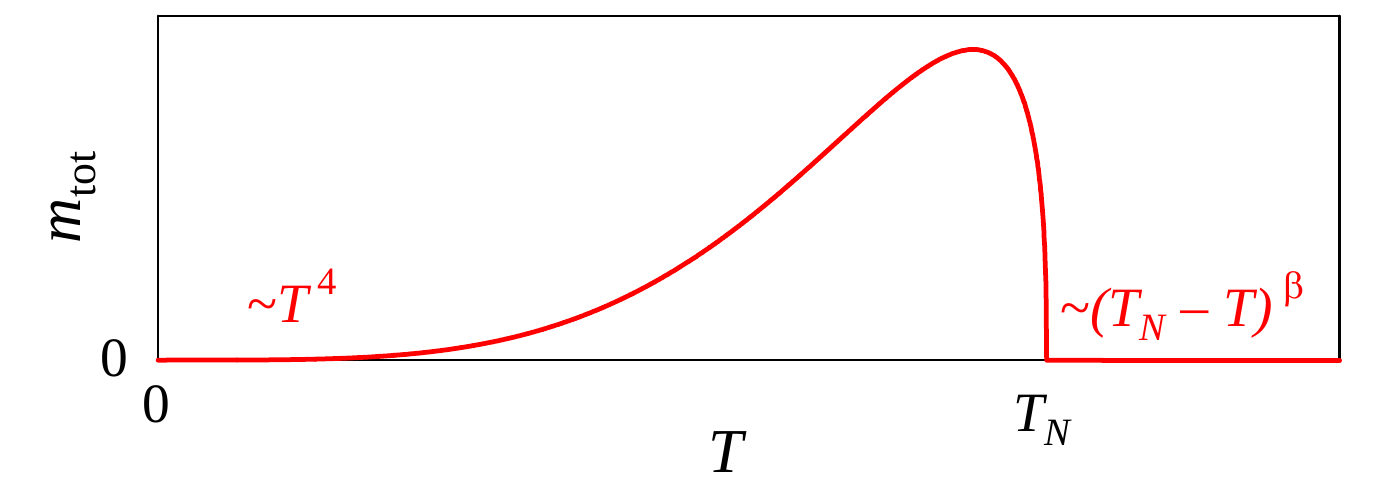}}
\caption{
Qualitative temperature dependence of the fluctuation-induced uniform magnetization in sublattice-imbalanced Heisenberg antiferromagnets, with the critical exponent $\beta=0.37$ \cite{janke93,campo02} in $d=3$ dimensions.
}
\label{fig:magschem}
\end{figure}

In this paper we discuss the phenomenon of fluctuation-induced ferrimagnetism in antiferromagnets, and we propose that {\scbo} at high pressure is in fact a ferrimagnet. Ferrimagnetism refers to states which display both staggered and uniform magnetizations, and it commonly occurs in systems with two different types of magnetic ions with unequal spin sizes \cite{wolf61}. Here, we identify a distinct mechanism for ferrimagnetism: In a system with equal-sized spins which displays N\'eel antiferromagnetism in the ground state, a uniform magnetization is induced at finite temperature solely by fluctuation effects.
More precisely, we show that thermal fluctuations generically produce a finite magnetization in a Heisenberg antiferromagnet once the $\Ztwo$ symmetry between the two sublattices is broken at the Hamiltonian level. Remarkably, quantum fluctuations do not produce a finite magnetization at $T=0$ due to spin conservation, such that the uniform magnetization becomes a non-monotonic function of temperature, as illustrated in Fig.~\ref{fig:magschem}.
We exemplify this in a layered toy model consisting of two interpenetrating square-lattice ferromagnets, for which we employ spin-wave theory to calculate the temperature-induced magnetization. 
In addition, simple Landau theory is used to analyze the behavior near the N\'eel temperature $\TN$.
We then consider a layered version of the Shastry-Sutherland model, as appropriate for the material \scbo. We predict the existence of a uniform magnetization in its orthorhombic high-pressure phase and provide a rough estimate for its amplitude.

The remainder of the paper is organized as follows: In Sec.~\ref{sec:toy} we introduce the toy model and demonstrate the phenomenon of fluctuation-induced ferrimagnetism. We also discuss temperature-induced corrections to spin-wave spectrum and link them to general hydrodynamics. Sec.~\ref{sec:scbo} is devoted to the layered Shastry-Sutherland model appropriate for {\scbo} where we provide quantitative results of relevance for its high-pressure phase. A discussion and outlook close the paper.


\section{Ferrimagnetism from thermal fluctuations: Toy model}
\label{sec:toy}

In this section we utilize a simple toy model to discuss the emergence of ferrimagnetism from thermal fluctuations in antiferromagnets with broken sublattice symmetry. We also connect the results to general aspects from Landau theory and hydrodynamics.

\subsection{Model}

\begin{figure}[t]
\centerline{\includegraphics[width=0.8\columnwidth]{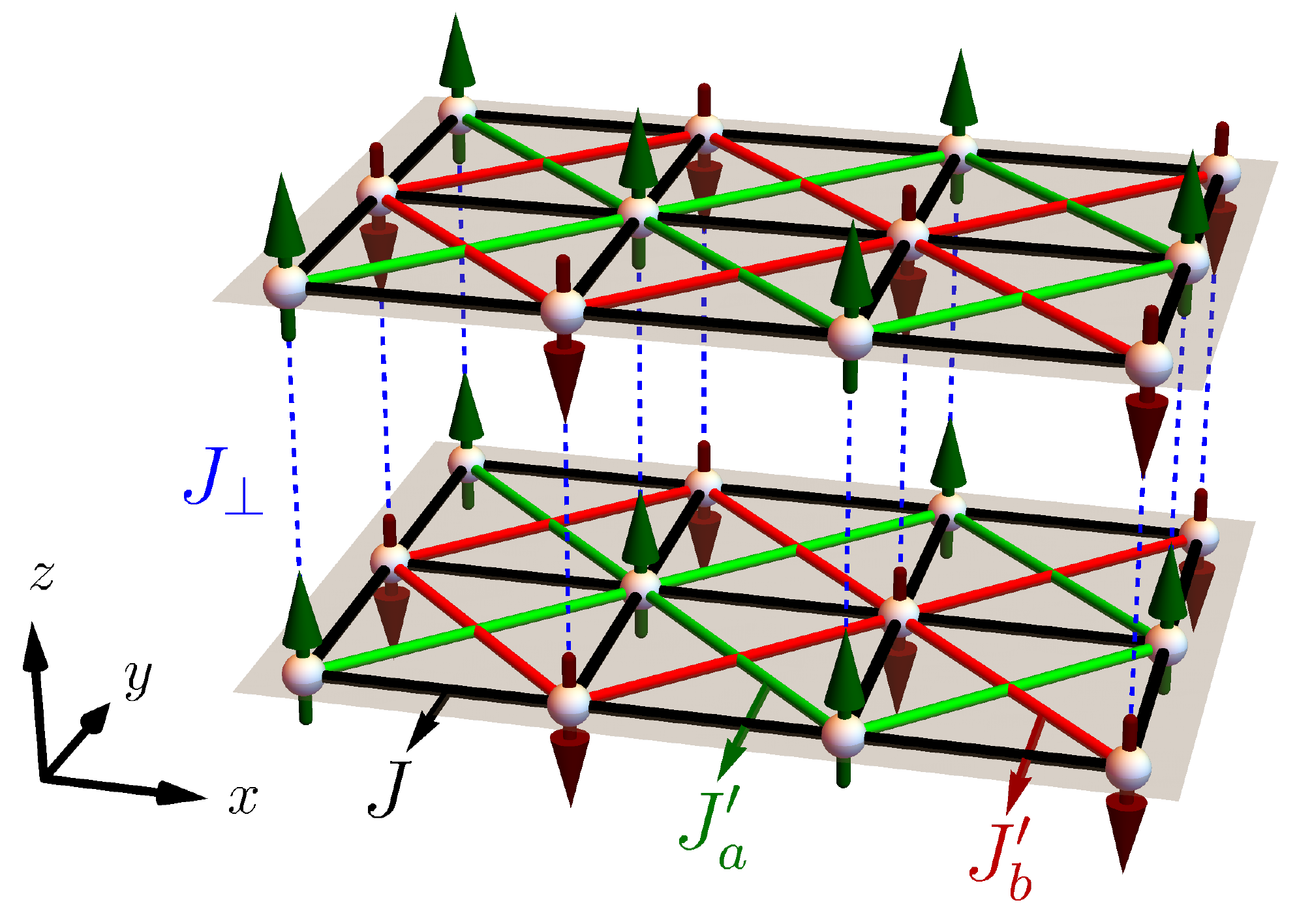}}
\caption{
Layered square-lattice Heisenberg antiferromagnet with two inequivalent second-neighbor couplings $J'_a$ (green) and $J'_b$ (red). Thermal fluctuations induce a finite uniform magnetization for $0<T<\TN$ once $J'_a \neq J'_b$.
}
\label{fig:toyschem}
\end{figure}

Our model is constructed from a bipartite square-lattice Heisenberg model with nearest-neighbor AF coupling $J$ between spins $S$, displaying collinear N\'eel order. The $\mathbb{Z}_2$ symmetry between the two sublattices is broken by adding second-neighbor couplings which are different for the two sublattices $A$ and $B$; we label them $J'_a$ and $J'_b$ and choose them to be ferromagnetic in order to stabilize N\'eel antiferromagnetism. Finally, we add a (small) ferromagnetic inter-layer interaction $J_\perp$ such that magnetic order also appears at finite temperatures. The model is depicted in Fig.~\ref{fig:toyschem}, its Hamiltonian reads
\begin{align}
\label{eq:htoy}
\mathcal{H} &=  J \sum_{\langle ij\rangle m} \vec S_{i,m} \cdot \vec S_{j,m}
- J_\perp \sum_{im} \vec S_{i,m} \cdot \vec S_{i,m+1} \notag \\
&- J'_a \!\! \sum_{\llangle ii'\in A\rrangle} \!\! \vec S_{i,m} \cdot \vec S_{i',m}
- J'_b \!\! \sum_{\llangle jj'\in B\rrangle} \!\! \vec S_{j,m} \cdot \vec S_{j',m}
\end{align}
where $i,j$ denote in-plane lattice coordinates, $m$ is the layer index, and $\langle ij\rangle$ and $\llangle ii'\rrangle$ denote pairs of first and second neighbors, respectively. The model displays a global $\SUtwo$ spin symmetry. For $J'_a\neq J'_b$ it features a two-site unit cell, and we will set the lattice constant of the underlying square lattice to unity.

Various limiting cases are of interest: On one hand, for $J\gg J'_{a,b}$ we have an antiferromagnet in which $J'_a\neq J'_b$ induces weak sublattice symmetry breaking. On the other hand, $J'_{a,b} \gg J$ corresponds to two inequivalent ferromagnets on the two sublattices which are weakly coupled by $J$ such that global collinear AF order emerges.

For purposes of illustration, we will also discuss the model with unequal spin sizes on the two sublattices, i.e., spins $S$ and $\eta S$ on sublattices $A$ and $B$, respectively. For $\eta\neq 1$, this then corresponds to the standard setting of a ferrimagnet. Unless noted otherwise, $\eta=1$ is assumed below.

We note that a {\em ferromagnetic} inter-layer coupling is crucial for the broken sublattice symmetry: If, instead, $J_\perp$ were antiferromagnetic, the long-range order would feature a two-layer periodicity, with $J'_a$ (as well as $J'_b$) acting on different sublattices in adjacent layers, such that, globally, the sublattice symmetry would be unbroken.

\subsection{Ground-state antiferromagnetism}
\label{sec:toygs}

In the classical limit, $S\to\infty$, the ground state of the model \eqref{eq:htoy} is obviously a two-sublattice collinear antiferromagnet, which has zero total magnetization $M$ and displays a global $\Uone$ symmetry. This symmetry implies that all quantum fluctuation processes occurring at finite $S$ on top of the collinear state conserve the longitudinal component of $M$. As a result, the magnetization remains zero at any order of a $1/S$ expansion, as will be demonstrated explicitly below.
In fact, this argument holds beyond the semiclassical limit: For infinitesimal $J$ the ground state consists of two saturated ferromagnets whose magnetizations are opposite, resulting in $M=0$. Since fluctuation processes on top of this state arise from $J$ only, $M$ remains zero at any order in perturbation theory in $J/J'$ for all $S$. From this we conclude that there is a stable AF quantum ground state which has zero total magnetization, protected by $\SUtwo$ symmetry, despite the broken sublattice symmetry.

We note that these arguments do not exclude the existence of phase transitions to other phases with non-zero $\mtot$ which might exist for finite $S$, strongly broken sublattice symmetry, and/or $J,J'$ of similar magnitude. However, for the toy model at hand such phases are not expected as long as the couplings $J'_{a,b}$ remain ferromagnetic.

\subsection{Spin-wave spectrum at $T=0$}
\label{sec:swspec}

\begin{figure}[t]
\centerline{\includegraphics[width=0.95\columnwidth]{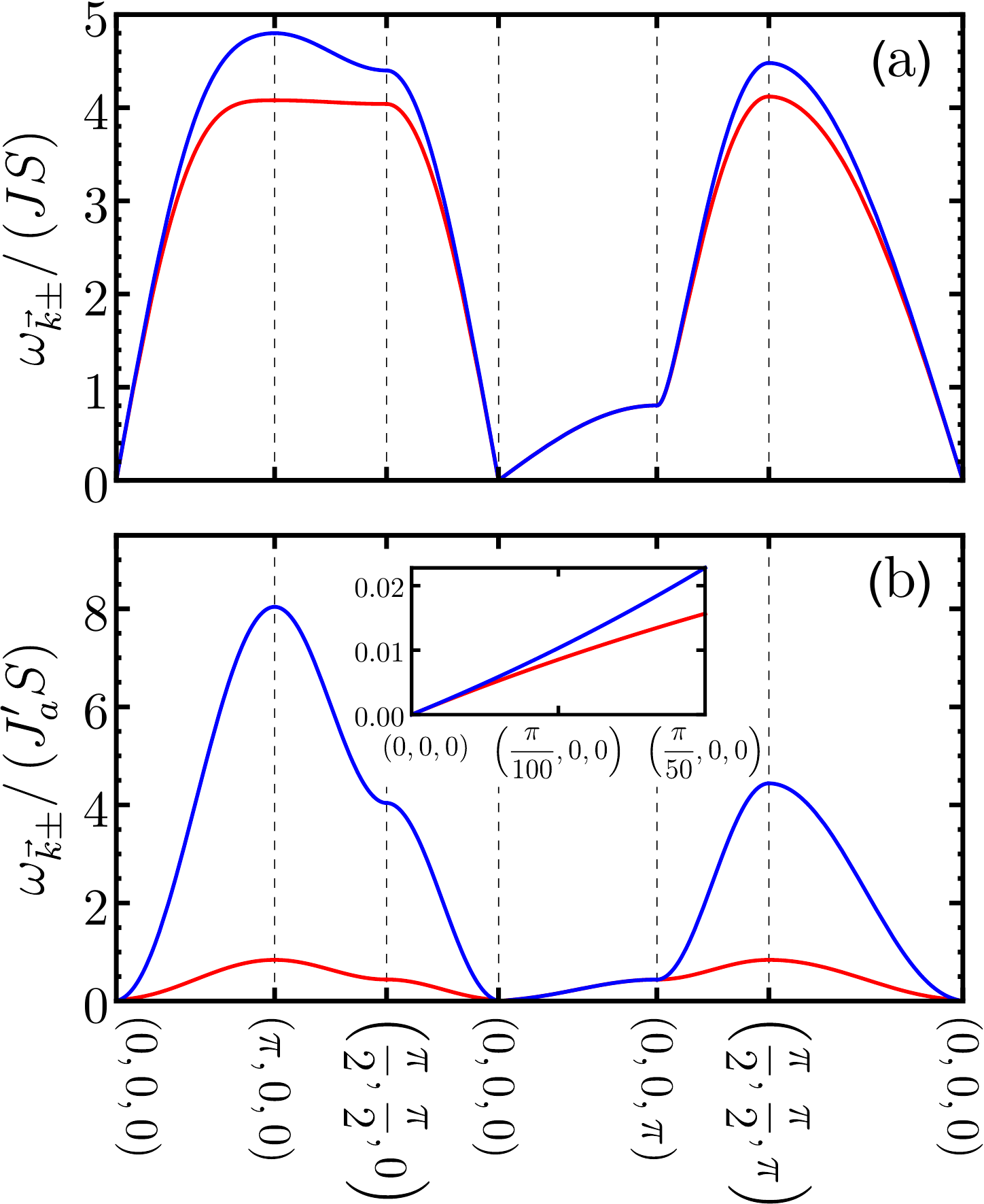}}
\caption{
Spin-wave dispersion for the toy model \eqref{eq:htoy} along a path in the Brillouin zone and parameters
(a) $J=100$, $J_a'=10$, $J_\perp=2$ and
(b) $J=0.1$, $J_a'=10$, $J_\perp=1$ in units of $J_b'$,
both with $\eta=1$.
Blue (red) curves correspond to $\w_{\vec{k}+}$ ($\w_{\vec{k}-}$), respectively. The inset in (b) shows a zoom into the low-energy part of the dispersion near $\vec{k}=\left(0,0,0\right)$.
}
\label{fig:toydisp}
\end{figure}

The broken sublattice symmetry has influence on the spin-wave spectrum. Since the ground state is a collinear antiferromagnet, we expect two Goldstone modes which are degenerate and linearly dispersing in the long-wavelength limit. This is confirmed by an explicit spin-wave calculation using the standard Holstein-Primakoff technique. The model features a two-site magnetic unit cell, such that all results can be obtained analytically, for details see Appendix \ref{app:toy}.
At leading order in $1/S$, i.e., by linear spin-wave theory, we obtain the dispersion of the two spin-wave modes as
\begin{equation}
\w_{\vec{k}\pm} = S \left(\sqrt{P_{\vec{k}}^2 - Q_{\vec{k}}^2} \pm R_{\vec{k}}\right)
\label{eq:toydisp1}
\end{equation}
with
\begin{align}
P_{\vec{k}} &= 2\left(J_a' + \eta J_b' \right) \xi_{\vec{k}} + \left(\eta+1\right) \left[2J +J_\perp \left(1-\cos{k_z}\right)\right],
\notag \\
Q_{\vec{k}} &= 4\eta^{1/2} J \gamma_{\vec{k}},
\notag \\
R_{\vec{k}} &= 2\left(J_a'-\eta J_b'\right)\xi_{\vec{k}} + \left(\eta-1\right)\left[2J -J_\perp \left(1-\cos{k_z}\right)\right],
\notag \\
\xi_{\vec{k}} &= 1 - \cos{k_x}\cos{k_y},
\notag \\
\gamma_{\vec{k}} &= \left( \cos{k_x} + \cos{k_y} \right)/2.
\end{align}
For $\eta=1$ we find two modes which are degenerate for $J'_a=J'_b$, but non-degenerate otherwise, i.e., the spectrum is split by broken sublattice symmetry. However, while the modes split at general $\vec{k}$, they have the same linear slope in the hydrodynamic limit of small $|\vec{k}|$. Denoting the in-plane wavevector as $\vec{k}_\parallel = (k_x,k_y)$ and $k_\parallel = |\vec{k}_\parallel|$, the mode energies can be expanded as
\begin{align}
\w_{\vec{k}\pm} & = 2 S \sqrt{2J} \sqrt{ \left(J_{a}' + J_{b}' + J \right) k_\parallel^2 + J_{\perp} k_z^2}
\notag \\ &
\pm S(J_a' - J_b') k_\parallel^2 + \mathcal{O} \left(k^3\right)
\label{eq:toydisp2}
\end{align}
for $J>0$. The mode dispersions are illustrated in Fig.~\ref{fig:toydisp} for parameter sets with (a) $J\!\gg\!J_{a,b}'$ and (b) $J\!\ll\!J_{a,b}'$. Note that $\w_{\vec{k}\pm}$ is symmetric with respect to rotations around the $k_z$ axis up to $\mathcal{O}\left(k^2\right)$. Moreover, the quadratic term in Eq.~\eqref{eq:toydisp2} vanishes when $J_a' = J_b'$, i.e., for equivalent sublattices. We also see a vanishing of the quadratic term for $k_\parallel=0$ because the interlayer coupling $J_\perp$ does not distinguish sublattice $A$ from $B$. Hence, spin waves with zero in-plane momentum do not experience the sublattice symmetry breaking.

It is instructive to discuss the limit $J'_{a,b} \gg J$, which, as mentioned earlier, describes two inequivalent and weakly coupled ferromagnetic subsystems. In this setting, decreasing $J$ should restrict the linear portion of the spectrum to smaller and smaller values of $k$ as the Goldstone modes approach the quadratic shape expected for decoupled ferromagnets. One can track this transformation by computing the nonzero wavenumber $k^\ast(\theta)$, with $\tan\theta=k_\parallel/k_z$, at which the magnitudes of the linear and quadratic terms in Eq.~\eqref{eq:toydisp2} become equal. A simple calculation yields
\begin{align}
k^\ast(\theta) &= \frac{2 \sqrt{2J}}{\left| J_a'-J_b' \right| \sin\theta } \left(J_a'+J_b'+J+\frac{J_\perp}{\tan^2\theta}\right)^{1/2},
\label{eq:kcrossover}
\end{align}
which confirms our expectation: For fixed $J_a' \ne J_b'$ and $J_\perp$, $k^*$ indeed decreases with $J$.

Inspecting the Bogoliubov coefficients, explicitly listed in Eq.~\eqref{eq:uvbogoliubov}, shows that the two modes have different weights on the two sublattices once $J_a' \ne J_b'$. More specifically, for $J'_a>J'_b$ the mode $+$ ($-$) is primarily located on sublattice $A$ ($B$).

We note that the low-energy behavior of the mode dispersion is qualitatively different for $\eta\neq 1$, and we will get back to this in Sec.~\ref{sec:hydro} below. This section will also discuss corrections to the mode dispersion beyond linear spin-wave theory.

\subsection{Uniform magnetization at low temperatures}

Spin-wave theory can be used to calculate fluctuation corrections to the sublattice magnetizations via a $1/S$ expansion. The next-to-leading order result, obtained from linear spin-wave theory (see Appendix \ref{app:toy} for details), reads
\begin{align}
m_A &= S - m_0 \left(T\right) + \frac{1}{N} \sum_{\vec{k}}\left(n_\mathrm{BE}^{\vec{k}-} - n_\mathrm{BE}^{\vec{k}+}\right),
\notag \\
m_B &= -\eta S + m_0 \left(T\right)+\frac{1}{N} \sum_{\vec{k}}\left(n_\mathrm{BE}^{\vec{k}-} - n_\mathrm{BE}^{\vec{k}+}\right).
\label{eq:mAmB}
\end{align}
Here, $n_\mathrm{BE}^{\vec{k}\pm} = 1/ (e^{\w_{\vec{k}\pm}/T}-1)$ is the Bose-Einstein distribution function where we have set Boltzmann's constant to unity, $k_B=1$, and
\begin{equation}
\label{eq:mzt}
m_0\left(T\right) = \sum_{\vec{k}} \frac{F_{\vec{k}}}{N}\left(1 + n_\mathrm{BE}^{\vec{k}-} + n_\mathrm{BE}^{\vec{k}+}\right) - \frac{1}{2},
\end{equation}
with $F_{\vec{k}}$ being a temperature-independent coefficient specified in Eq.~\eqref{eq:Fk}. One can see that the quantum corrections, $m_0(T=0)$, are equal on both sublattices, leading to vanishing total magnetization at $T=0$ for $\eta=1$, as announced in Sec.~\ref{sec:toygs} above.
The thermal corrections, however, are different because the non-degenerate spin-wave modes experience different thermal occupations. As a result, we obtain a uniform magnetization per site,
\begin{align}
\mtot = \frac{\left(1-\eta\right)}{2} S + \frac{1}{N} \sum_{\vec{k}}\left(n_\mathrm{BE}^{\vec{k}-} - n_\mathrm{BE}^{\vec{k}+}\right),
\label{eq:toymag}
\end{align}
which is finite for non-zero temperature even if the spin sizes on the two sublattices are equal, $\eta=1$. This is a central result of this paper.

We expect these expressions to yield reliable results for low temperatures where the occupation of the spin-waves modes remains small to validate the approximation of noninteracting magnons. This corresponds to having small $1/S$ corrections in Eq.~\eqref{eq:mAmB} in comparison to the leading-order term.

\begin{figure}[t]
\centerline{\includegraphics[width=0.9\columnwidth]{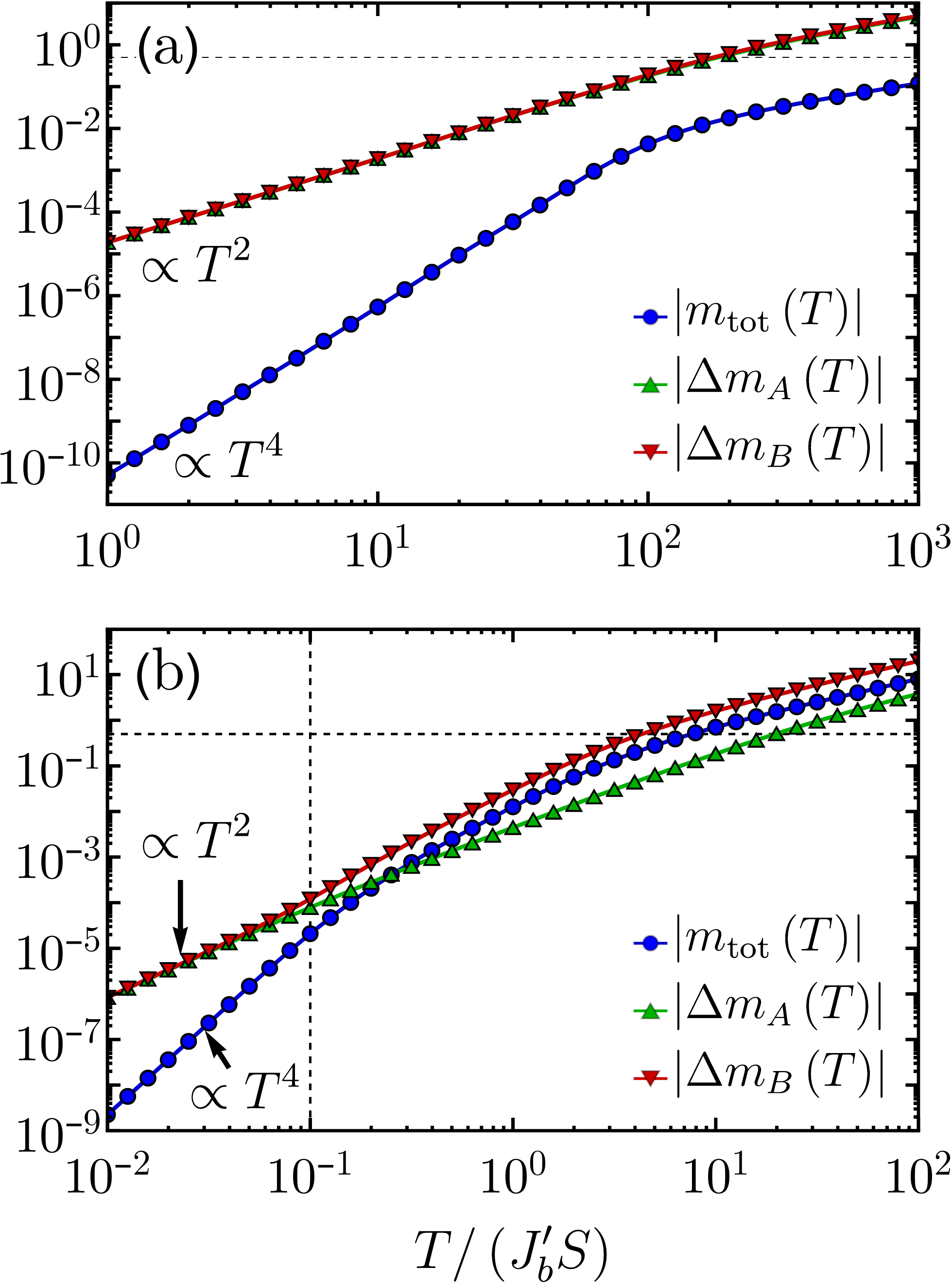}}
\caption{
Fluctuation-induced uniform magnetization, $\mtot$, as function of temperature, together with the finite-temperature corrections to the sublattice magnetizations, $\Delta m_{A,B} = m_{A,B}(T)-m_{A,B}(0)$. The coupling constants were set to
(a) $J=100$, $J_a'=10$, $J_\perp=2$ and
(b) $J=0.1$, $J_a'=10$, $J_\perp=1$
in units of $J_b'$. In both plots, the horizontal grid line marks the value $1/2$, whereas the vertical grid line in (b) indicates the temperature $T=JS$.
}
\label{fig:toymag1}
\end{figure}

The fluctuation-induced uniform magnetization emerges at order $S^0$ in the spin-wave expansion.
Fig.~\ref{fig:toymag1} depicts its temperature dependence, together with the thermal corrections to the sublattice magnetizations, at a fixed ratio $J_a'/J_b'=10$ when (a) $J\!\gg\!J_{a,b}'$ and (b) $J\!\ll\!J_{a,b}'$. In both cases, the low-temperature corrections to $m_A$ and $m_B$ scale as $T^2$, a feature which is also observed in antiferromagnets without broken sublattice symmetry \cite{oguchi60}. This can be easily rationalized by power counting: For linearly dispersing modes, $F_{\vec{k}}$ scales as $1/k$, and hence the leading contribution to the $T$ dependence of the integral in Eq.~\eqref{eq:mzt} has the form $\int dk k^{d-2} n_\mathrm{BE}(k/T)$ and scales as $T^{d-1}$.

In this low-temperature regime, the only accessible excited states lie within the energy range where the magnon branches are nearly degenerate, so that a difference in the occupation of the spin-wave modes emerges as a subleading effect in $T$. Expanding the dispersions in next-to-leading order in $k$ and noting that the diverging $F_{\vec{k}}$ factor is absent from Eq.~\eqref{eq:toymag}, we find that the uniform magnetization scales as $T^4$, see Appendix \ref{app:toy}.

In Fig.~\ref{fig:toymag1}(a), we see that for $J\!\gg\!J_{a,b}'$ this low-$T$ power law continues well beyond the point where $T/S$ matches the smallest coupling in the system, $J_b'$, and extends up to temperatures $T \sim JS$, beyond which spin-wave theory is no longer valid. The reason is that strong $J$ yields both large spin-wave velocities at small $k$ \eqref{eq:toydisp2} \textit{and} large values of $k^\ast(\theta)$ \eqref{eq:kcrossover}, such that a significant difference in the occupations of the spin-wave modes only appears at relatively high energies. As a result, $\mtot$ only reaches values of $10^{-3}$ at the temperature where thermal corrections to $m_{A,B}$ become significant, i.e., of order $10^{-1}$.
The opposite limit, $J\!\ll\!J_{a,b}'$, leads to a markedly different behavior, shown in Fig.~\ref{fig:toymag1}(b). The now weak coupling $J$ stabilizes approximately sublattice-symmetric AF order only at very low temperatures. Thus, once $T\!\gtrsim\!JS$, sublattice $B$ becomes much more susceptible to fluctuations than sublattice $A$, which explains why $m_B$ ($m_A$) starts growing faster (slower) than $T^2$ around the vertical dashed line. The deviation from the low-temperature scaling is then responsible for enhancing the uniform magnetization, which becomes as large as $5\cdot10^{-2}$ when thermal corrections to $m_B$ reach $10^{-1}$.

\begin{figure}[t]
\centerline{\includegraphics[width=0.9\columnwidth]{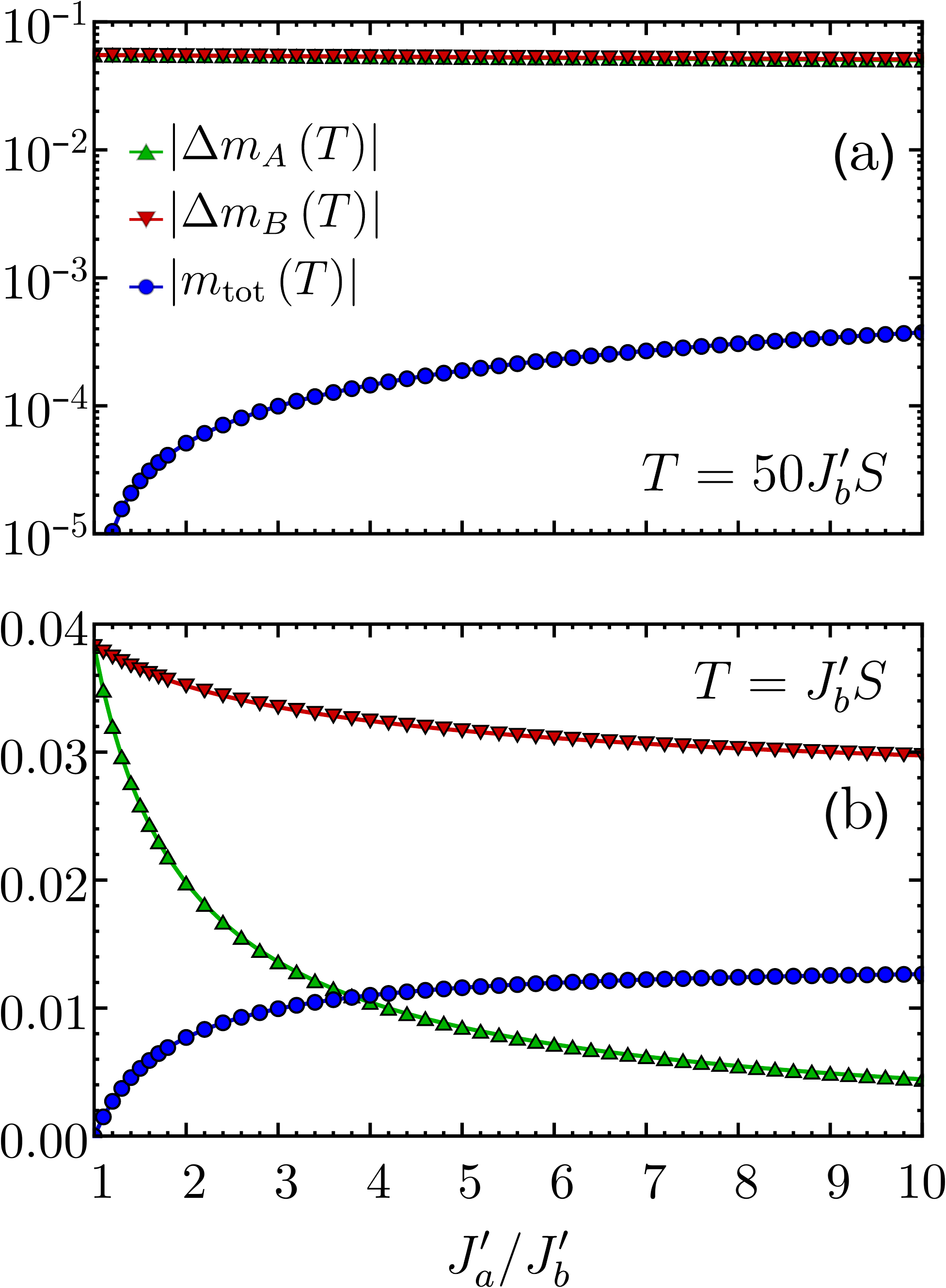}}
\caption{
Fluctuation-induced magnetization as in Fig.~\ref{fig:toymag1}, but now as function of $J'_a/J'_b$ at fixed finite temperature. Parameters were set to (a) $T/S=50$, $J=100$, $J_\perp=2$ and (b) $T/S=1$, $J=0.1$, $J_\perp=1$ all in units of $J_b'$, and are such that the results with the ratio $J_a'/J_b'=10$ correspond to data in Fig.~\ref{fig:toymag1} at the specified temperatures.
}
\label{fig:toymag2}
\end{figure}

Fig.~\ref{fig:toymag2} illustrates how the same quantities in Fig.~\ref{fig:toymag1} vary with the ratio $J_a'/J_b'$ at fixed $T/(J_b'S)$ when (a) $J\!\gg\!J_{a,b}'$ and (b) $J\!\ll\!J_{a,b}'$. The uniform magnetization vanishes in the limit $J_a'/J_b'\to1$, which corresponds to restoring sublattice symmetry, while $\mtot \propto (J_a'/J_b'-1)$ for small imbalance. However, the increase in $\mtot$ saturates at a point where $J_a'/J_b'$ becomes so large that the spin-wave velocities in Eq.~\eqref{eq:toydisp2} start growing at the same rate that $k^*\left(\theta\right)$ decreases in Eq.~\eqref{eq:kcrossover}. Put more simply, saturation occurs when increasing $J_a'/J_b'$ only leads to further splitting of the spin-wave bands at energies larger than $T/S$.
The main difference between Figs.~\ref{fig:toymag2}(a) and (b) lies in the order of magnitude of the fluctuation-induced effects: At $J_a'/J_b'=10$, $\mtot$ is two orders of magnitude larger in panel (b) compared to panel (a). Once again, this difference follows from the fact that the strong AF coupling $J$ between the sublattices in (a) balances how fluctuations act on each of them, therefore diminishing the resulting uniform magnetization.

\subsection{Uniform magnetization at elevated temperatures}

To access elevated temperatures where spin-wave theory is no longer reliable, we resort to arguments from Landau theory. The sublattice-imbalanced antiferromagnet -- equivalent to a ferrimagnet -- is described by two order parameters, a staggered magnetization and a uniform magnetization, which are linearly coupled \cite{neumann94}.
As a result, there is a single transition upon cooling at $\TN$ from the paramagnet to the ferrimagnet, and both order parameters are zero (non-zero) above (below) $\TN$, respectively. The linear coupling also implies that the onset of the uniform magnetization below $\TN$ is identical to that of the staggered magnetization, hence $\mtot \propto (\TN-T)^\beta$ where $\beta$ is the order-parameter exponent \cite{stanley_rmp}. For the classical phase transition at hand, the universality class for the Heisenberg magnet remains O(3) independent of whether the transition is into an antiferromagnetic or ferrimagnetic state. Hence, $\beta=0.37$ in $d=3$ space dimensions \cite{janke93,campo02}.
Together with the low-temperature result $\mtot\propto T^4$, we conclude that the uniform magnetization displays a non-monotonic temperature dependence as illustrated in Fig.~\ref{fig:magschem}.

Again, it is instructive to discuss the limit $J\ll J'_{a,b}$. For $J'_a > J'_b$, the transition at $\TN$ concerns primarily the onset of ferromagnetism on the A sublattice. Weak $J$ produces a small opposite magnetization on the $B$ sublattice, resulting in a collinear ferrimagnet. (Note that the energy gained by maintaining the collinearity of the global magnetic order is extensive, whereas the entropy associated with directional fluctuations of sublattice $B$ is only intensive.) In the limit of large sublattice imbalance, $J'_a \gg J, J'_b$, there is hence a window of temperatures below $\TN$ where the uniform magnetization is large, $\mtot\approx m_A/2$ as $m_B\ll m_A$.
In this limit, it is also easy to see that the sign of the uniform magnetization is the same at low $T$ and close to $\TN$: It is the sublattice with weaker magnetism that experiences stronger thermal fluctuations, such that $\mtot$ aligns with the magnetization of the more strongly ordered sublattice, i.e., the $A$ sublattice if $J'_a > J'_b$.

\subsection{Hydrodynamic modes and corrections to the spin-wave spectrum}
\label{sec:hydro}

For a broader picture, we connect our findings to general hydrodynamic considerations. A collinear two-sublattice antiferromagnet, spontaneously breaking $\SUtwo$ symmetry, is characterized by a non-conserved order parameter and displays two linearly dispersing Goldstone modes which are degenerate in the long-wavelength limit. This is in agreement with the spin-wave result \eqref{eq:toydisp2} for $\eta=1$.

A ferrimagnet, in contrast, has in addition a conserved order parameter, namely uniform magnetization $\mtot$. As a result, it features a single quadratically dispersing Goldstone mode
\cite{kaplan58,rademaker20}. This can be nicely seen in the explicit spin-wave expressions \eqref{eq:toydisp1} for $\eta\neq 1$: Here $\w_{\vec{k}-}$ is gapless and quadratic in $k$ whereas $\w_{\vec{k}+}$, though also quadratic, exhibits a gap given by $4\left|\eta-1\right|JS$.

Together, this implies that the low-energy spectrum of the model \eqref{eq:htoy} must change qualitatively when going from $T=0$ to $T>0$: The system turns from an antiferromagnet to a ferrimagnet, such that one of the $T=0$ Goldstone modes must acquire a temperature-induced gap, and the other one must change its dispersion from linear to quadratic at small $k$. This change can be captured by non-linear spin-wave theory, i.e., has the form of $1/S$ corrections at finite $T$. While it is straightforward to write down the quartic terms in the spin-wave Hamiltonian, analyzing all terms at finite temperature turns out to be rather laborious, and therefore we refrain from doing so. However, as these corrections are suppressed as $T\to 0$, they have no influence on the leading low-temperature behavior of the uniform magnetization, $\mtot\propto T^4$.


\section{Ferrimagnetism in a layered distorted Shastry-Sutherland model}
\label{sec:scbo}

After having established that sublattice-imbalanced antiferromagnets generically display fluctuation-induced ferrimagnetism, we now turn to an experimentally relevant example, namely the Shastry-Sutherland lattice as realized in the compound \scbo.

\subsection{Model and symmetries}

\begin{figure}[t]
\centerline{\includegraphics[width=0.85\columnwidth]{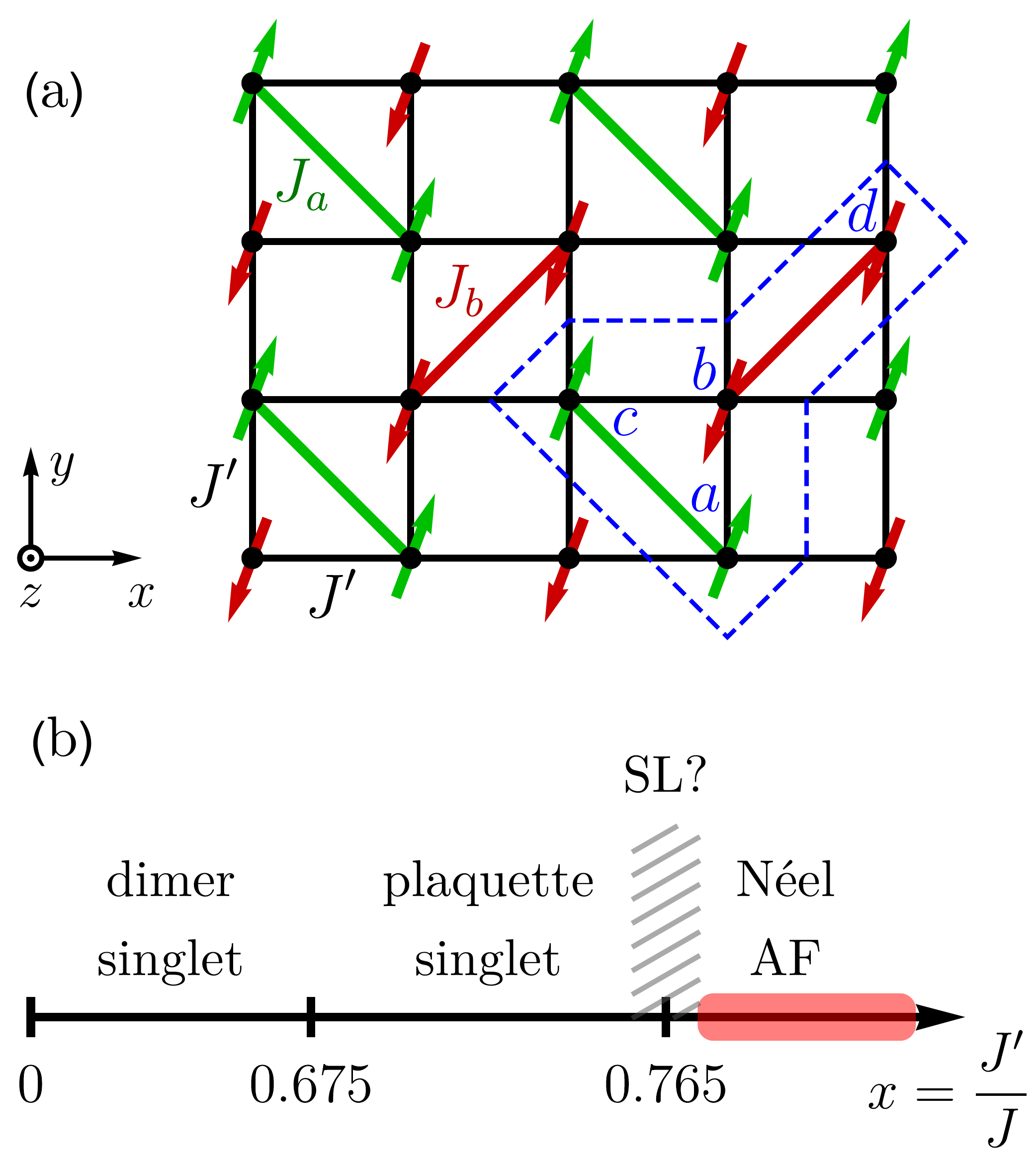}}
\caption{
(a) Shastry-Sutherland model with intra-dimer couplings $J_{a,b}$ and inter-dimer coupling $J'$. The dashed lines indicate the unit cell.
(b) Ground-state phase diagram of the $S=1/2$ Shastry-Sutherland model with $J_a=J_b$, as reported in Ref.~\onlinecite{corboz13}. An intermediate spin-liquid (SL) phase (shaded) \cite{foot1} has been recently proposed in Ref.~\onlinecite{yang21}. In the present work, the focus is on the antiferromagnetic phase at large $J'/J$ shown in red.
}
\label{fig:scboplane}
\end{figure}

Our starting point is the Heisenberg model on the Shastry-Sutherland lattice, consisting of orthogonal dimers of spins $1/2$ with intra-dimer coupling $J$ and inter-dimer coupling $J'$, Fig.~\ref{fig:scboplane}(a). This model features a four-site unit cell, containing two dimers, and displays, in addition to mirror symmetries along the dimer axes, a non-symmorphic glide symmetry which maps the two types of dimers into each other.

The phase diagram of the Shastry-Sutherland model has been determined numerically \cite{koga00,corboz13}, Fig.~\ref{fig:scboplane}(b): It contains a paramagnetic dimer phase for $x=J'/J<0.675$, a bipartite N\'eel antiferromagnet for $x>0.765$, and a plaquette-ordered singlet paramagnet, the so-called empty-plaquette phase, in between \cite{corboz13}. In addition, a very recent numerical study \cite{yang21} has proposed that a gapless quantum spin-liquid phase is realized in a narrow range, $0.79<x<0.82$, intervening between the plaquette-singlet and AF phases \cite{foot1}.

\begin{figure}[t]
\centerline{\includegraphics[width=0.9\columnwidth]{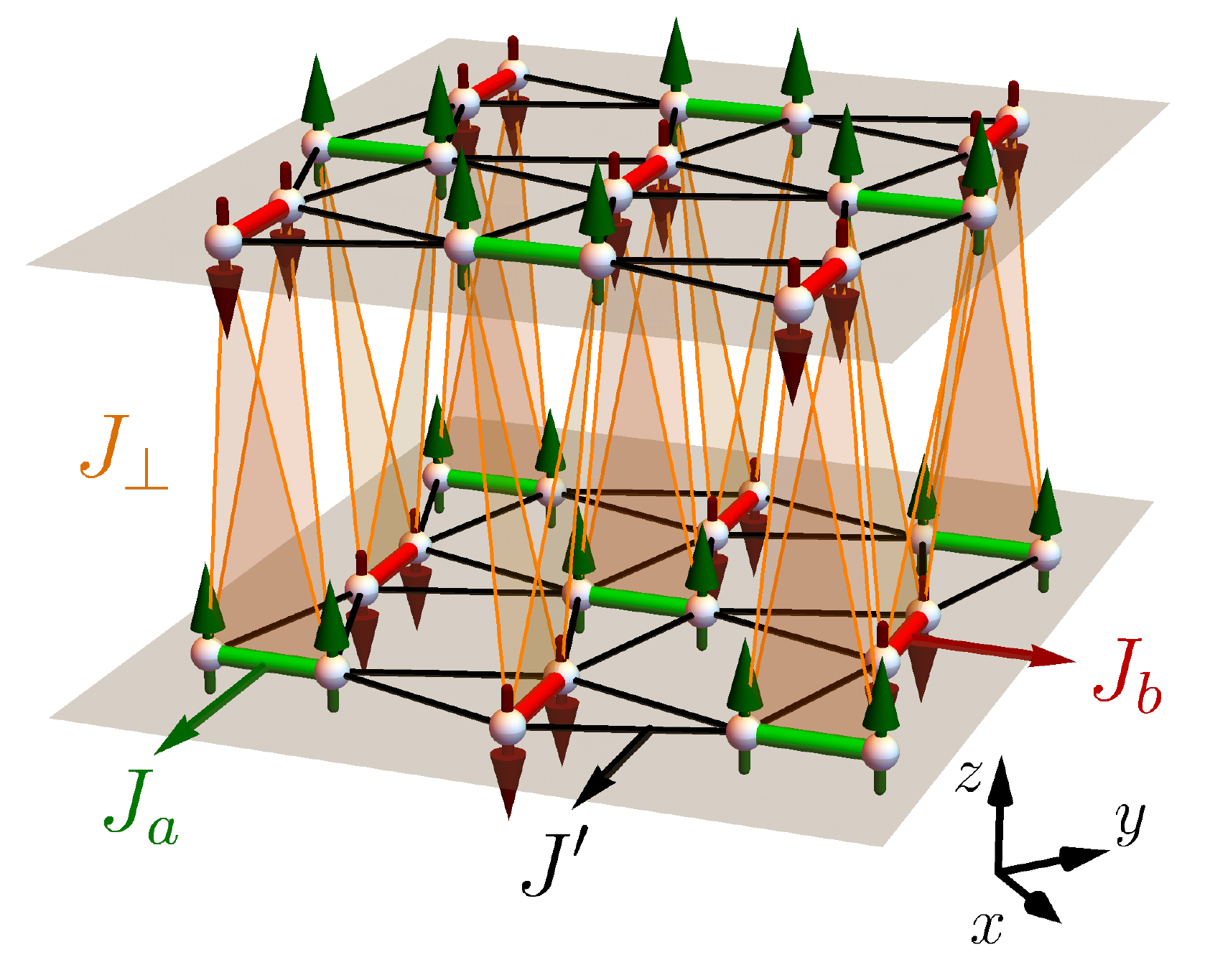}}
\caption{
Layered Shastry-Sutherland model \eqref{eq:hss} used to describe \scbo. An orthorhombic distortion is assumed to generate different intra-dimer couplings $J_a$, $J_b$.
}
\label{fig:scboschem}
\end{figure}

The ferrimagnetism discussed in this paper appears upon breaking the glide symmetry, such that two different types of (mutually parallel) dimers emerge. Such symmetry breaking corresponds to an orthorhombic distortion where half of the intra-dimer bonds elongate and the other half contract \cite{moliner11,boos19}. In the AF state, each of the intra-dimer couplings acts on one AF sublattice only, such that the symmetry between the two sublattices is broken.
In the following we will therefore consider a distorted Shastry-Sutherland model with intra-dimer couplings $J_{a,b}$ and inter-dimer coupling $J'$. To meaningfully discuss magnetic order at finite temperature, we work with a layered version of the model. Guided by the structure of \scbo, we consider a stacking of the layers such that orthogonal dimers are on top of each other, and include a (small) antiferromagnetic Heisenberg interlayer coupling $J_\perp$ which pairwise connects vertically stacked dimers \cite{miya00}. The model, illustrated in Fig.~\ref{fig:scboschem}, is described by the Hamiltonian
\begin{align}
\label{eq:hss}
\mathcal{H} &=
J_a \!\!\sum_{\llangle ij \in A\rrangle m} \!\! \vec S_{i,m} \cdot \vec S_{j,m}
+ J_b \!\! \sum_{\llangle ij \in B\rrangle m} \!\! \vec S_{i,m} \cdot \vec S_{j,m} \notag \\
& + J' \sum_{\langle ij\rangle} \vec S_{i,m} \cdot \vec S_{j,m} \notag \\
& + J_\perp \sum_{\llangle ij\rrangle m} (\vec S_{i,m}+\vec S_{j,m}) \cdot (\vec S_{i,m+1} + \vec S_{j,m+1})
\end{align}
where each term in the last sum represents four couplings between the spins of neighboring dimers in $z$ direction, and we consider spins of general size $S$.

The ground states of the single-layer version of the distorted Shastry-Sutherland model \eqref{eq:hss} with $S=1/2$ have been studied in Refs.~\onlinecite{moliner11,boos19}. While all phases of the original Shastry-Sutherland model appear stable against a small dimer imbalance, the main finding of Ref.~\onlinecite{moliner11} is the existence of a Haldane-like phase for strongly imbalanced dimers and weak inter-dimer coupling $J'$. This phase is dominated by one-dimensional correlations; it is adiabatically connected to a so-called full-plaquette phase and has been argued \cite{boos19} to be a candidate for the intermediate phase observed experimentally in \scbo.

\subsection{Spin-wave theory in the antiferromagnetic phase}

As announced, we are interested in antiferromagnets with broken sublattice symmetry. Hence we focus on the physics of the model \eqref{eq:hss} in the regime of larger $x=J'/J$, where one encounters a clear connection to the toy model discussed in Sec.~\ref{sec:toy}: Both systems display a collinear antiferromagnetic classical ground state with two sublattices, each of which experiences an independent internal coupling. Therefore, the two models share the same mechanism for breaking sublattice symmetry.

As in Sec.~\ref{sec:toy}, we perform a spin-wave calculation to determine its properties in a $1/S$ expansion both at zero and finite temperature. For the standard 2D Shastry-Sutherland model, the linear-spin-wave theory description of the AF phase breaks down for $x<1$, signaling a transition to a different phase at this level of the approximation. Hence, we work with parameter sets corresponding to $x\gtrsim 1$.

In the AF phase of model \eqref{eq:hss}, the symmetry-broken state features four sites per unit cell, such that the Bogoliubov transformation can only be performed numerically. For convenience, we employ an in-plane coordinate system corresponding to the square lattice shown in Fig.~\ref{fig:scboplane}(a), such that the basis vectors of the (magnetic) unit cell are given by $a_1=(2a,0,0)$, $a_2=(0,2a,0)$, and $a_3=(a,a,c)$ where $a$ and $c$ are the in-plane and out-of-plane lattice constants which we set to unity in the following. The relevant details of the calculation are given in Appendix~\ref{app:ss}; here we summarize the key results.

\begin{figure}[t]
\centering
\includegraphics[width=\columnwidth]{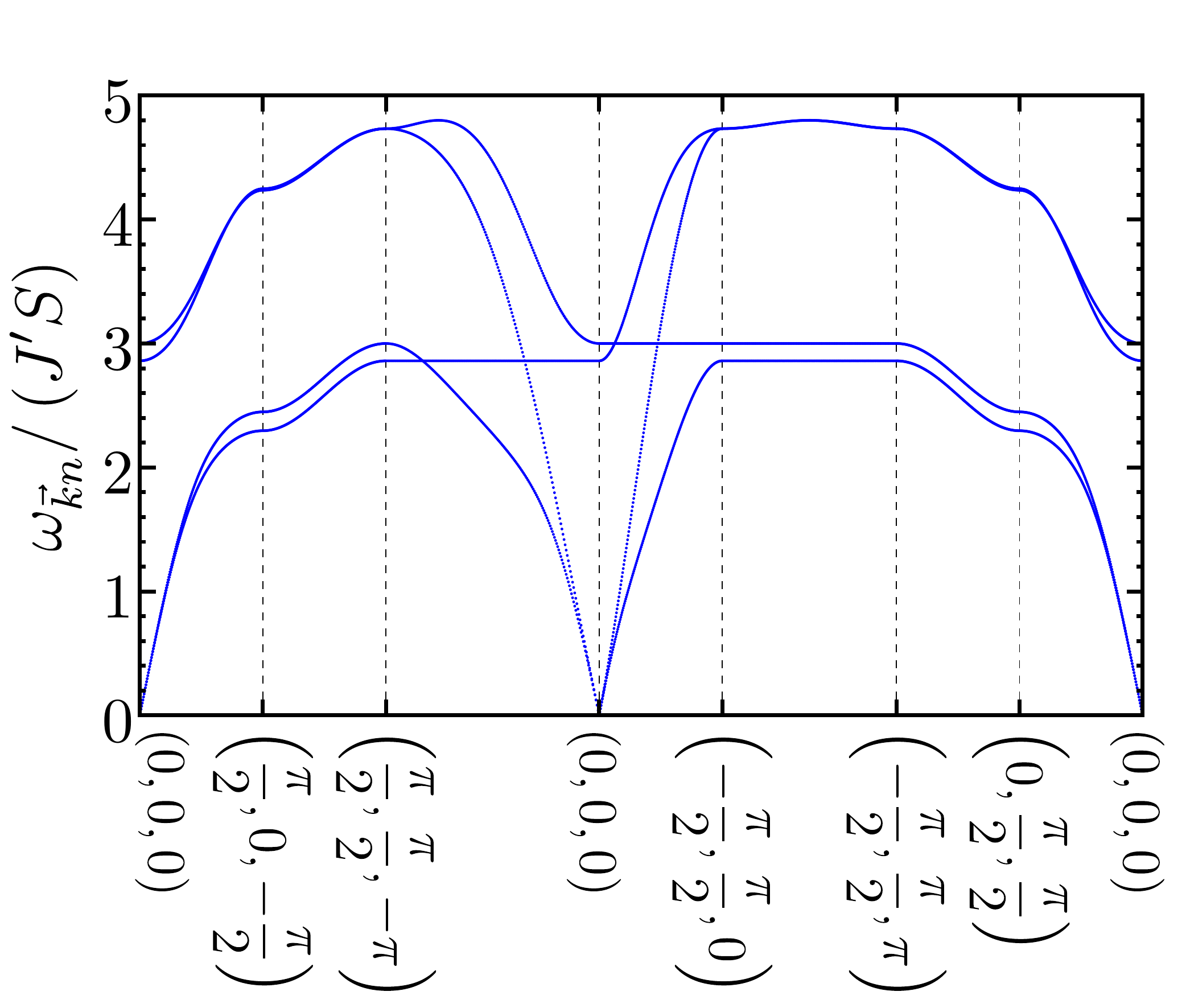}
\caption{
Spin-wave dispersion of the distorted Shastry-Sutherland model \eqref{eq:hss} along a path in the Brillouin zone for parameters $J_a=0.97$, $J_b=0.9$ and $J_\perp=0.2$ in units of $J^\prime$.
}
\label{fig:ssdisp}
\end{figure}

The spin-wave spectrum along an exemplary path in the BZ is illustrated in Fig.~\ref{fig:ssdisp}. Of the four spin-wave modes, two are gapped at small momenta, while the two others are linearly dispersing Goldstone modes. As with the toy model, these modes are degenerate only for $J_a = J_b$, i.e. when the sublattice symmetry is preserved; for $J_a\neq J_b$ they share the same velocity, but differ at quadratic order except for $k_\parallel=0$. Notably, the Goldstone-mode velocity is highly anisotropic, e.g., it is different even for different in-plane directions because $J_a\neq J_b$ leaves only mirror symmetries intact.

\begin{figure}[t]
\centerline{\includegraphics[width=0.9\columnwidth]{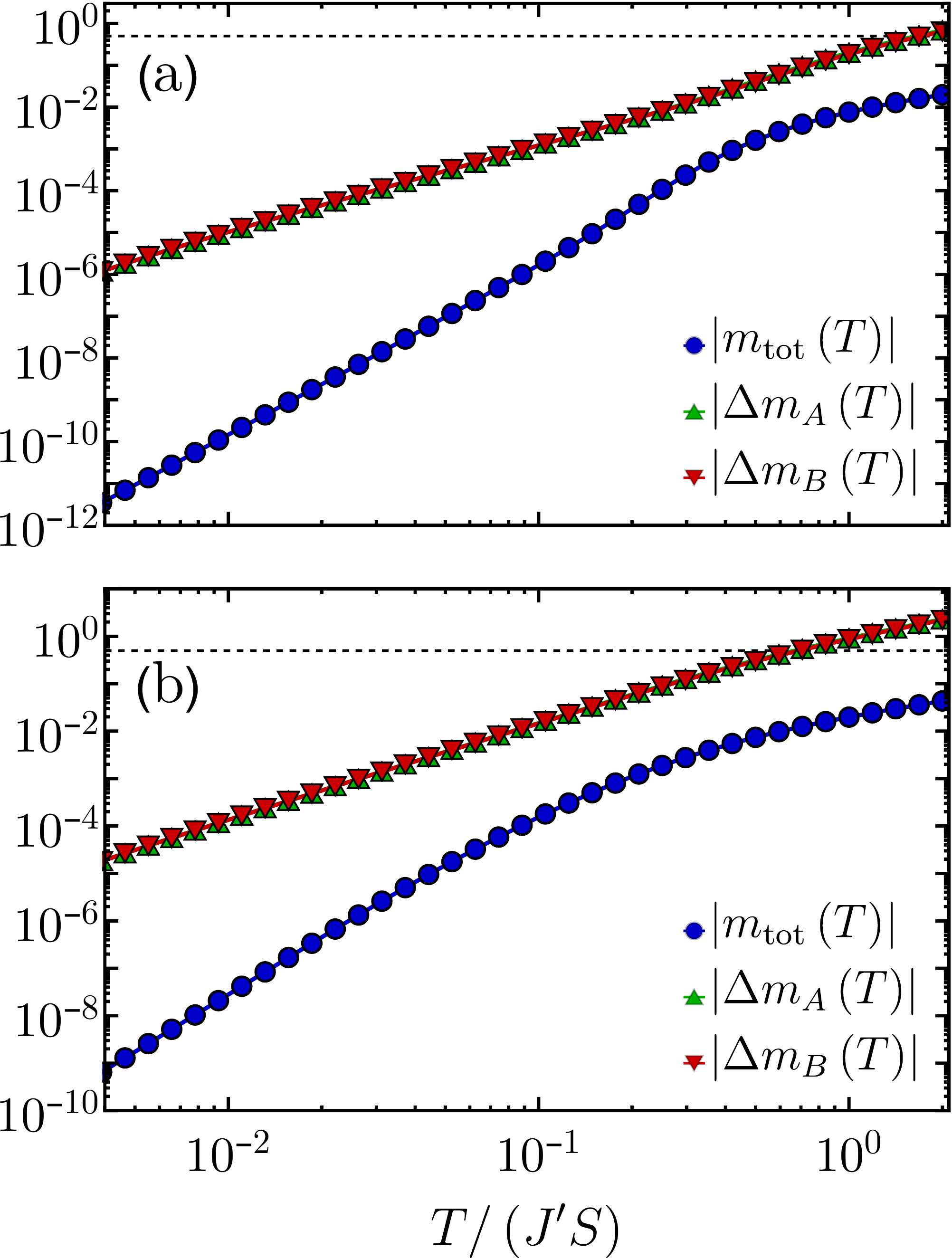}}
\caption{
Uniform magnetization, $\mtot$, and thermal corrections, $\Delta m_{A,B}$, to the sublattice magnetization of the distorted Shastry-Sutherland model, with parameters
(a) $J_a=0.9$, $J_b=0.8$, $J_\perp=0.1$ and
(b) $J_a=0.97$, $J_b=0.9$, $J_\perp=0.01$ in units of $J'$.
}
\label{fig:ssmag}
\end{figure}

\subsection{Ferrimagnetism}

The qualitative arguments for fluctuation-induced ferrimagnetism brought forward in Sec.~\ref{sec:toy} apply unchanged to the sublattice-imbalanced antiferromagnet of the Shastry-Sutherland model.
Our numerical evaluation of $1/S$ corrections to the magnetizations on the individual sites of the unit cell, as detailed in Appendix~\ref{app:ss}, confirms this expectation. The quantum corrections are equal on all sites, resulting in a vanishing total magnetization at $T=0$.
In contrast, the thermal corrections are different on the $A$ (up) and $B$ (down) sublattices, while they are pairwise equal on the two unit-cell sites belonging to the $A$ and $B$ sublattice, respectively. As before, the thermal corrections to the sublattice magnetizations, $\Delta m_A(T)$ and $\Delta m_B(T)$, scale proportional to $T^2$ at low temperature. The uniform magnetization, $\mtot = [m_A(T) + m_B(T)]/2$, scales as $T^4$ because two mode dispersions differ at quadratic order only.

Numerical results illustrating the variation of the magnetization with temperature are shown in Fig.~\ref{fig:ssmag}. The parameters in panels (a) and (b) correspond to weaker (stronger) fluctuation corrections, driven both by the different $J_\perp$ and by $J_{a,b}$ being further away (closer) to the critical value $J_{a,b}=J'$ within spin-wave theory. Consequently, the uniform magnetization is much larger in (b) compared to (a) at the same temperature, even though the ratio $J_a/J_b$ is similar in both cases. While the extreme limit of two weakly coupled ferromagnets, discussed for the toy model, cannot be realized in the Shastry-Sutherland model, the magnetization nevertheless can get as large as $5\cdot 10^{-3}$ at the temperature where the largest $\Delta m$ is $10^{-1}$.

\begin{figure}[t]
	\centerline{\includegraphics[width=0.9\columnwidth]{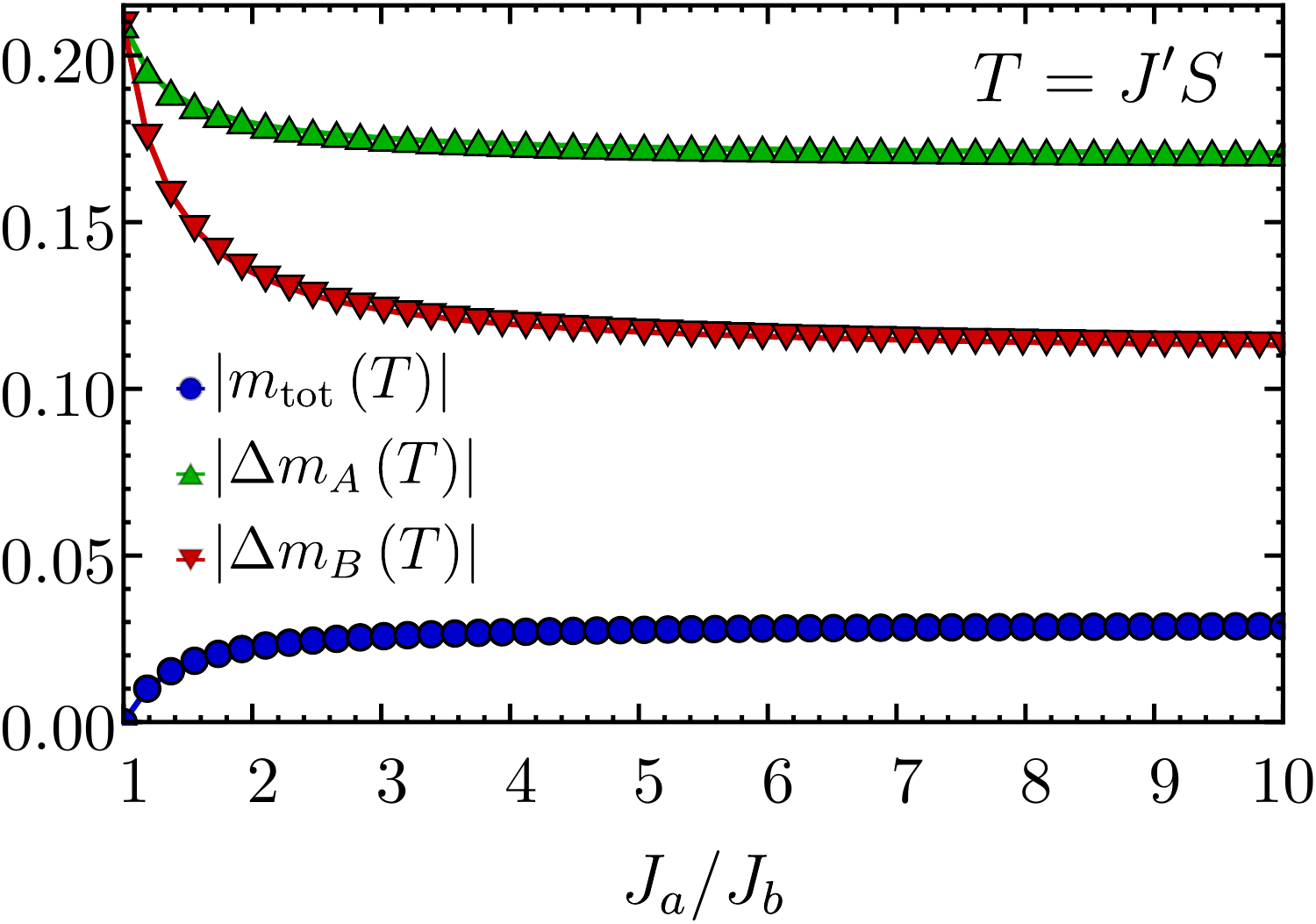}}
	\caption{
Same quantities as in Fig.~\ref{fig:ssmag}, but now as a function of $J_a/J_b$ at fixed temperature $T=J'S$ and with $J_a=0.9$ (and varying $J_b$) and $J_\perp=0.1$ in units of $J'$. The ratio $J_a/J_b = 1.125$ reproduces the data in Fig.~\ref{fig:ssmag}(a) at the specified temperature.
}
	\label{fig:ssmagjb}
\end{figure}

Fig.~\ref{fig:ssmagjb} depicts the effect of varying the sublattice imbalance at a fixed temperature while keeping $J'$ as the largest coupling in the system. Differently from the toy model, the thermal corrections are now larger on the strongly coupled sublattice, since the AF couplings $J_{a,b}$ are frustrated. Still, the uniform magnetization shows the same trend as in Fig.~\ref{fig:toymag2}, growing with increasing sublattice imbalance until it saturates at large $J_a/J_b$. For {\scbo}, small orthorhombic distortion likely implies that $J_a/J_b$ remains close to unity.

\subsection{Application to S\lowercase{r}C\lowercase{u}(BO$_3$)$_2$ under pressure}

{\scbo} assumes a tetragonal structure at ambient pressure and low temperatures, where its magnetic properties are in very good agreement with those of the two-dimensional Shastry-Sutherland model in the small-$x$ dimer phase. The magnetic couplings have been estimated to be $J\approx 85$\,K and $J'\approx54$\,K \cite{miya99,miya00}.

High-pressure studies of {\scbo} detected various signatures of pressure-driven phase transitions. In particular, indications for a different, but still paramagnetic, phase were found above $2$\,GPa \cite{waki07}, and this transition was later located more precisely to be around $1.8$\,GPa \cite{hara12,zayed14,zayed17,sakurai18,guo20,jimenez21}. While it is natural to assume that this paramagnetic phase represents the empty-plaquette phase of the Shastry-Sutherland model, both the NMR results of Ref.~\onlinecite{waki07} and the neutron scattering results of Ref.~\onlinecite{zayed17} appear to be incompatible with this idea: The empty-plaquette phase displays equivalent magnetic sites and $C_4$ symmetry while the NMR data indicate the existence of two inequivalent magnetic sites. To resolve this contradiction, it has been argued \cite{boos19} that an orthorhombic distortion, stabilizing a different plaquette phase in the intermediate regime, is most compatible with the NMR \cite{waki07} and neutron scattering \cite{zayed17} data.

At higher pressures, a structural transition to a monoclinic structure occurs around $4.5$\,GPa \cite{hara12}, and AF order with a rather high N\'eel temperature of $120$\,K has been detected at $5.5$\,GPa via neutron scattering \cite{hara14}. It has been suggested, but not clarified beyond doubt, that this magnetic order in fact emerges around $4$\,GPa before the structural transition \cite{zayed15}.
%
In addition, a recent low-temperature thermodynamic study \cite{guo20} found indications for a previously undetected AF state below $4$\,K occurring between $3$ and $4.2$\,GPa. It has been suggested \cite{guo20} that it is this low-temperature AF state which should be interpreted as the genuine AF state of the Shastry-Sutherland model, given that the higher-pressure monoclinic system no longer features the orthogonal dimers characteristic of the Shastry-Sutherland model.
The same study also presented evidence for an additional phase transition occurring above $4.2$\,GP at $8$\,K and proposed that this transition is related to the existence of yet another low-temperature magnetic state, which is likely to display AF order as well. However, a full characterization of this phase is still lacking.
Apparently, more work is needed to discern the fascinating high-pressure physics of \scbo.

For our purpose, we focus on the fact that pressure-induced structural distortions lead to inequivalent dimers; this likely applies to all pressures larger than $2$\,GPa \cite{waki07,hara14,boos19}. AF order in each layer will thus be sublattice-imbalanced because of the broken glide symmetry. Achieving a finite uniform magnetization then relies on the uniform magnetization in adjacent layers being parallel.
Our model in Fig.~\ref{fig:scboschem} assumes the established layer stacking, with orthogonal dimers on top of each other \cite{miya00,hara14}, an antiferromagnetic interlayer coupling \cite{miya00,hara14,guo20}, and an orthorhombic distortion of the tetragonal structure as proposed in Ref.~\onlinecite{boos19}. Together, this yields a macroscopic uniform magnetization, which we can estimate to reach up to $5\cdot10^{-3} \mu_B$ per Cu atom at its temperature maximum, i.e., slightly below the N\'eel temperature, see Fig.~\ref{fig:magschem}. It would be very interesting to test this prediction in future high-pressure magnetization measurements.
For such an experiment, one needs to keep in mind that ferrimagnets generically form magnetization domains much like ferromagnets \cite{wolf61}; detecting a uniform magnetization might therefore require cooling in an applied field.

We note that the type of dimer distortion suggested to exist at $5.5$\,GPa at elevated $T$ in Ref.~\onlinecite{hara14}, see their Figs. 5 and 6, would lead to a vanishing total magnetization instead, because a strong up-spin dimer in one layer would couple to a strong (instead of weak) down-spin dimer in the next layer. The resulting finite magnetization per layer may still be detectable as a surface effect.

Finally, we comment on the effect of small Dzyaloshinskii-Moriya (DM) interactions which break $\SUtwo$ symmetry at the Hamiltonian level and are known to be present in {\scbo} \cite{cepas2001,hara14}. In the AF phase of interest here, DM interactions may lead to a small, but finite, uniform magnetization at $T=0$ and also alter its leading low-temperature corrections as the spin-wave spectrum may acquire a gap at $T=0$. Nonetheless, the conclusion that inequivalent sublattices experience different thermal fluctuations remains, and for small DM interactions the non-monotonic temperature dependence of the magnetization will be preserved.


\section{Conclusions and outlook}
\label{sec:concl}

In summary, we have shown that collinear N\'eel antiferromagnets, whose $\Ztwo$ symmetry between the two sublattices is broken in the Hamiltonian, become ferrimagnets at finite temperature. Interestingly, this is an effect driven by thermal fluctuations but not by quantum fluctuations, constituting an interesting example where both types of fluctuations produce different physics  -- in contrast to many instances of order by disorder where thermal and quantum fluctuations lead to very similar state selection \cite{villain80}.
We have proposed that such a fluctuation-induced ferrimagnetic phase is realized in {\scbo} under high pressure, where a lattice distortion breaks the glide symmetry of the Shastry-Sutherland lattice.

Our work suggests a number of future directions: First, it is conceivable that similar thermal fluctuation effects occur in antiferromagnets with more complicated ground-state spin structures. Second, while we have argued that the antiferromagnetic phase with $\mtot=0$ is a stable state of matter at $T=0$ despite sublattice imbalance, it is interesting to ask whether quantum fluctuations can generate additional, more non-trivial, zero-temperature phases in sublattice-imbalanced antiferromagnets. Finally, considering the same phenomenology in the presence of charge carriers will lead to a fluctuation-induced anomalous Hall effect.


\acknowledgments

We thank L. Janssen  and E. Andrade for discussions and collaborations on related work.
Financial support from the Deutsche Forschungsgemeinschaft through SFB 1143 (project-id 247310070) and the W\"urzburg-Dresden Cluster of Excellence on Complexity and Topology in Quantum Matter -- \textit{ct.qmat} (EXC 2147, project-id 390858490) is gratefully acknowledged.


\appendix

\section{Spin-wave calculations for toy model}
\label{app:toy}

The starting point for our analysis of the toy model proposed in Sec.~\ref{sec:toy} was to expand the Hamiltonian in Eq.~\eqref{eq:htoy} in powers of $1/\sqrt{S}$ around a collinear N\'eel state. Since we allow the two magnetic sublattices to have, at least in principle, spins of unequal sizes, the Holstein-Primakoff transformation reads
\begin{align}
S_i^z &= S - a_i^\dagger a_i,
\notag \\
S_i^+ &= \sqrt{2S-a_i^\dagger a_i} \, a_i
= \sqrt{2S} \,a_i + \mathcal{O} \left(1/\sqrt{S}\right),
\notag \\
S_i^- &= a_i^{\dagger} \sqrt{2S-a_i^\dagger a_i}
= \sqrt{2S} \, a_i^\dagger + \mathcal{O} \left(1/\sqrt{S}\right),
\label{eq:HPsubA}
\end{align}
for sites $i$ located on sublattice $A$ and
\begin{align}
S_i^z &= -\eta S + b_i^\dagger b_i
\notag \\
S_i^+ &= b_i^{\dagger} \sqrt{2\eta S-b_i^\dagger b_i}
= \sqrt{2\eta S} \, b_i^\dagger + \mathcal{O} \left(1/\sqrt{\eta S}\right),
\notag \\
S_i^- &= \sqrt{2\eta S-b_i^\dagger b_i} \, b_i
= \sqrt{2\eta S} \, b_i + \mathcal{O} \left(1/\sqrt{\eta S}\right),
\label{eq:HPsubB}
\end{align}
for sites belonging to sublattice $B$. As usual, $a_i^\dagger$ and $b_i^\dagger$ are bosonic creation operators with corresponding annihilation operators $a_i$ and $b_i$.

After substituting Eqs.~\eqref{eq:HPsubA} and \eqref{eq:HPsubB} into the Hamiltonian, applying a Fourier transform to the bosonic operators, and neglecting terms beyond $\mathcal{O} \left(S\right)$, one arrives at a quadratic Hamiltonian of the form
\begin{equation}
\mathcal{H}_\mathrm{LSW} = S^2 E_\mathrm{cl} + \frac{S}{2}\sum_{\vec{k}} \left[\Psi_{\vec{k}}^{\dagger}\mathbb{M}_{\vec{k}}\Psi_{\vec{k}} - \left(A_{\vec{k}}+B_{\vec{k}}\right) \right],
\label{eq:toyHLSW}
\end{equation}
where
\begin{equation}
S^2 E_\mathrm{cl} = -NS^2 \left[ 2\eta J + J_{a}' + \eta^{2}J_{b}' + \frac{\left(1+\eta^{2}\right)}{2}J_{\perp}\right]
\end{equation}
is the classical ground-state energy for a system with $N$ sites, $\Psi_{\vec{k}} = \left(a_{\vec{k}},b_{\vec{k}},a_{-\vec{k}}^\dagger,b_{-\vec{k}}^\dagger \right)^\mathrm{T}$, and
\begin{equation}
\mathbb{M}_{\vec{k}}
=
\begin{pmatrix}
A_{\vec{k}} & 0 & 0 & C_{\vec{k}}\\
0 & B_{\vec{k}} & C_{\vec{k}} & 0\\
0 & C_{\vec{k}} & A_{\vec{k}} & 0\\
C_{\vec{k}} & 0 & 0 & B_{\vec{k}}
\end{pmatrix}
\end{equation}
with
\begin{align}
A_{\vec{k}} &= 4 \left(\eta J + J_a'\xi_{\vec{k}}\right) + 2 J_{\perp}\left(1 - \cos k_z \right),
\notag \\
B_{\vec{k}} &= 4 \left(J+\eta J_b'\xi_{\vec{k}}\right)+2\eta J_{\perp}\left(1-\cos k_z\right),
\notag \\
C_{\vec{k}} &= 4\eta^{1/2} J \gamma_{\vec{k}}.
\label{eq:abc}
\end{align}

The linear spin-wave Hamiltonian in Eq.~\eqref{eq:toyHLSW} can be diagonalized by means of a Bogoliubov transformation
\begin{align}
a_{\vec{k}} &= u_{\vec{k}} \alpha_{\vec{k}} - v_{\vec{k}} \beta_{-\vec{k}}^\dagger,
& 
b_{\vec{k}} &= u_{\vec{k}} \beta_{\vec{k}} - v_{\vec{k}} \alpha_{-\vec{k}}^\dagger
\end{align}
with coefficients
\begin{align}
u_{\vec{k}} &= \mathrm{sgn}\left(C_{\vec{k}}\right) \sqrt{\frac{F_{\vec{k}}+1}{2}},
&
v_{\vec{k}} &= \sqrt{\frac{F_{\vec{k}}-1}{2}}
\label{eq:uvbogoliubov}
\end{align}
given in terms of the function
\begin{equation}
F_{\vec{k}} = \frac{A_{\vec{k}} + B_{\vec{k}}}{ \sqrt{\left(A_{\vec{k}} + B_{\vec{k}}\right)^2 - 4 C_{\vec{k}}^2} }.
\label{eq:Fk}
\end{equation}
When expressed in terms of the Bogoliubov operators, Eq.~\eqref{eq:toyHLSW} takes on the diagonal form
\begin{align}
\mathcal{H}_\mathrm{LSW} &= S^2 E_\mathrm{cl} + \frac{1}{2} \sum_{\vec{k}} \left[ \w_{\vec{k}-} + \w_{\vec{k}+} - S\left( A_{\vec{k}}+B_{\vec{k}} \right)\right]
\notag \\
& + \sum_{\vec{k}} \left( \w_{\vec{k}+} \alpha_{\vec{k}}^\dagger \alpha_{\vec{k}} + \w_{\vec{k}-} \beta_{\vec{k}}^\dagger \beta_{\vec{k}}\right).
\label{eq:htoylsw}
\end{align}
The resulting mode dispersions $\w_{\vec{k}\pm}$ are specified in Eq.~\eqref{eq:toydisp2} of the main text, with the abbreviations being related to the terms defined in Eq.~\eqref{eq:abc} as $P_{\vec{k}} = (A_{\vec{k}} + B_{\vec{k}})/2$, $R_{\vec{k}} = (A_{\vec{k}} - B_{\vec{k}})/2$, and $Q_{\vec{k}} = C_{\vec{k}}$.

We can then use the previous relations to derive the sublattice magnetizations to next-to-leading order in $1/S$. For sublattice $A$, we have
\begin{align}
m_A &= \frac{2}{N} \sum_{i\in A} \left\langle S_i^z \right\rangle
= S - \frac{2}{N} \sum_{\vec{k}} \left\langle a_{\vec{k}}^\dagger a_{\vec{k}} \right\rangle
\notag \\
&= S - \frac{2}{N} \sum_{\vec{k}} \left( u_{\vec{k}}^2 \left\langle \alpha_{\vec{k}}^\dagger \alpha_{\vec{k}} \right\rangle + v_{\vec{k}}^2 \left\langle 1 + \beta_{\vec{k}}^\dagger \beta_{\vec{k}} \right\rangle \right)
\notag \\
&= S - \frac{1}{N} \sum_{\vec{k}} \left[  \left( F_{\vec{k}}+1 \right) n_\mathrm{BE}^{\vec{k}+} + \left(F_{\vec{k}}-1\right)\left(1+n_\mathrm{BE}^{\vec{k}-}\right) \right],
\end{align}
which is equal to the first expression in Eq.~\eqref{eq:mAmB}. One recovers the second expression straightforwardly after a similar sequence of steps for sublattice $B$.

The low-temperature behavior of the uniform magnetization can in fact be predicted analytically by expanding the spin-wave dispersion up to quadratic order in $k$ and retaining the leading contribution in $T$ to Eq.~\eqref{eq:toymag}. Let us focus on the case $\eta=1$ for concreteness. After rewriting Eq.~\eqref{eq:toydisp2} in the form
\begin{equation}
\omega_{\vec{k}\pm} \approx \omega_1 \left(\theta\right) k \pm \omega_2 \left(\theta\right) k^2,
\end{equation}
we find
\begin{align}
n_\mathrm{BE}^{\vec{k}-} - n_\mathrm{BE}^{\vec{k}+} &\approx \frac{2e^{-\beta\omega_1 k} \sinh{\left(\beta\omega_2 k^2\right)}}{1-2e^{-\beta\omega_1 k}\cosh{\left(\beta\omega_2 k^2\right)}+e^{-2\beta\omega_1 k}}
\notag \\
& \approx 2 e^{-\beta \omega_1 k} \beta \omega_2 k^2.
\label{eq:nkdiff}
\end{align}
The approximation made in the last step of \eqref{eq:nkdiff} is justified by the fact that we are dealing with momenta $k\!\lesssim\! 1/\left(\beta \omega_1\right)$, and hence $\beta \omega_2 k^2 \!\lesssim \omega_2/\left(\beta \omega_1^2 \right) \!\ll\! 1$. We then substitute Eq.~\eqref{eq:nkdiff} into Eq.~\eqref{eq:toymag} to obtain
\begin{align}
\mtot &\propto \beta \int_0^\pi \! d\theta \, \sin\theta \, \omega_2\!\left(\theta\right) \int_0^\infty \! dk \, k^4 e^{-\beta \omega_1\!\left(\theta\right) k}
\notag \\
&\propto \frac{\left|J_a'-J_b'\right|}{\left(\beta S\right)^4 J^5}  \int_0^\pi \! d\theta  \, \sin\theta \, \left[ \frac{\omega_2\!\left(\theta\right)}{\left|J_a'-J_b'\right|S} \right] \, \left[\frac{JS}{\omega_1\!\left(\theta\right)}\right]^5\!.
\end{align}
Since the integral in the last line above is expressed solely in terms of dimensionless quantities, we conclude that in the low-temperature limit
\begin{equation}
\mtot \propto \frac{\left|J_a'-J_b'\right|}{J}\left(\frac{T}{JS}\right)^4,
\end{equation}
which is precisely what our numerical results indicate.


\section{Spin-wave calculations for Shastry-Sutherland model}
\label{app:ss}

The spin-wave calculations for the Shastry-Sutherland model are based on a Holstein-Primakoff representation of the spin operators as above. Since the unit cell consists of four spins which exhibit collinear N\'eel order in the classical limit, we assign the transformations given by Eqs.~\eqref{eq:HPsubA} and \eqref{eq:HPsubB} to two spins each, with spin size $S$ on all sublattices. Here, the spins on sublattice $A$ ($B$) correspond to Holstein-Primakoff operators $a_i$ and $c_i$ ($b_i$ and $d_i$), respectively, see Fig.~\ref{fig:scboplane}(a).

As noted in the main text, for calculation purposes we choose a coordinate system where the Shastry-Sutherland plane is located on a square lattice along the $J^\prime$ bonds. Moreover, the planes are stacked such that inequivalent dimers are positioned on top of each other. With lattice constants set to unity, this yields real-space basis vectors $a_1=(2,0,0)$, $a_2=(0,2,0)$, $a_3=(1,1,1)$ and reciprocal-space basis vectors $b_1=\pi(1,0,-1)$, $b_2=\pi(0,1,-1)$, $b_3=2\pi(0,0,1)$.

Substitution of the Holstein-Primakoff operators into the Hamiltonian given by Eq.~\eqref{eq:hss} and a subsequent Fourier transform yields the quadratic Hamiltonian
\begin{equation}
\mathcal{H}_\mathrm{LSW} = S^2 E_\mathrm{cl} + \frac{S}{2} \sum_{\vec{k}} \left[\Psi_{\vec{k}}^{\dagger}\mathbb{M}_{\vec{k}}\Psi_{\vec{k}} - 2 B_{\vec{k}} \right]
\label{eq:ssHLSW}
\end{equation}
where
\begin{equation}
S^2 E_\mathrm{cl} = N S^2 \left[\frac{J_a + J_b}{4} - 2J^\prime - 2J_\perp \right]
\end{equation}
is again the classical ground state energy for a system consisting of $N$ spins, $\Psi_{\vec{k}} = \left(a_{\vec{k}},c_{\vec{k}},b_{-\vec{k}}^\dagger,d_{-\vec{k}}^\dagger \right)^\mathrm{T}$ and
\begin{equation}
\mathbb{M}_{\vec{k}}
=
\begin{pmatrix}
A_{\vec{k}} & C_{\vec{k}} & E_{\vec{k}} & F_{\vec{k}}\\
C^\ast_{\vec{k}} & A_{\vec{k}} & F^\ast_{\vec{k}} & E^\ast_{\vec{k}}\\
E^\ast_{\vec{k}} & F_{\vec{k}} & B_{\vec{k}} & D_{\vec{k}}\\
F^\ast_{\vec{k}} & E_{\vec{k}} & D^\ast_{\vec{k}} & B_{\vec{k}}
\end{pmatrix}
\end{equation}
where $^\ast$ denotes the complex conjugate and
\begin{align}
A_{\vec{k}} &= 4 J^\prime - J_a + 4 J_\perp, \notag \\
B_{\vec{k}} &= 4 J^\prime - J_b + 4 J_\perp, \notag \\
C_{\vec{k}} &= J_a e^{i(-k_x+k_y)}, \notag \\
D_{\vec{k}} &= J_b e^{i(k_x+k_y)}, \notag \\
E_{\vec{k}} &= 2J^\prime \cos k_y + J_\perp ( e^{i(-k_x+k_z)} + e^{i (-k_x-k_z)} ), \notag \\
F_{\vec{k}} &= 2J^\prime \cos k_x + J_\perp ( e^{i( k_y+k_z)} + e^{i ( k_y-k_z)} ).
\end{align}

The Hamiltonian in Eq.~\eqref{eq:ssHLSW} can now be diagonalized by means of a generalized Bogoliubov transformation \cite{wessel05}
\begin{equation}
\Psi_{\vec{k}} = T(\vec{k}) \Phi_{\vec{k}}.
\label{eq:ssBT}
\end{equation}
with $\Phi_{\vec{k}} = \left(\alpha_{\vec{k}},\gamma_{\vec{k}},\beta_{\vec{k}}^\dagger,\delta_{\vec{k}}^\dagger \right)^\mathrm{T}$ being the normal mode spinor. The columns of $T(\vec{k})$ correspond to the eigenvectors of $\Sigma \mathbb{M}_{\vec{k}}$ with
\begin{equation}
\Sigma =
\begin{pmatrix}
\mathbb{I} & 0\\
0 & -\mathbb{I}
\end{pmatrix}
\end{equation}
where $\mathbb{I}$ denotes the $2\times 2$ unit matrix. Solving the eigenvalue problem for $\Sigma \mathbb{M}_{\vec{k}}$ yields two positive eigenvalues $\lambda_{\vec{k}1,2}$  and two negative eigenvalues $\lambda_{\vec{k}3,4}$. The normal modes of the Hamiltonian are then given by
\begin{align}
\omega_{\vec{k}1,2} &= S\lambda_{\vec{k}1,2}, \notag \\
\omega_{\vec{k}3,4} &= -S\lambda_{\vec{k}3,4}.
\label{eq:ssEvals}
\end{align}
Importantly, the positive and negative eigenvalues are \emph{not} of pairwise equal magnitude.
With Eqs.~\eqref{eq:ssBT} and \eqref{eq:ssEvals} the linear spin-wave Hamiltonian then takes the diagonal form
\begin{align}
\mathcal{H}_\mathrm{LSW} &= S^2 E_\mathrm{cl} + \sum_{\vec{k}} \left[\frac{\omega_{\vec{k}3} + \omega_{\vec{k}4}}{2} - S B_{\vec{k}} \right] \notag \\
&+ \sum_{\vec{k}} \left[\omega_{\vec{k}1} \numop{\alpha_{\vec{k}}} + \omega_{\vec{k}2} \numop{\gamma_{\vec{k}}} + \omega_{\vec{k}3} \numop{\beta_{\vec{k}}} + \omega_{\vec{k}4} \numop{\delta_{\vec{k}}} \right].
\end{align}

The sublattice magnetization can then be derived from the Holstein-Primakoff operators using the respective Bogoliubov coefficients and Eq.~\eqref{eq:ssBT}. The magnetizations on the $a$ and $c$ sites, both belonging to sublattice $A$, are equal and read
\begin{align}
m_A
&= \frac{2}{N} \sum_{i\in A} \expval{S_i^z} = S - \frac{4}{N} \sum_{\vec{k}} \expval{a_{\vec{k}}^{\dagger} a_{\vec{k}}} \notag \\
&= S - \frac{4}{N} \sum_{\vec{k}} \Big(
\absval{T_{11}(\vec{k})}^2 \expval{\numop{\alpha_{\vec{k}}}}
+ \absval{T_{12}(\vec{k})}^2 \expval{\numop{\gamma_{\vec{k}}}} \notag \\
&+ \absval{T_{13}(\vec{k})}^2 \expval{1+\numop{\beta_{\vec{k}}}}
+ \absval{T_{14}(\vec{k})}^2 \expval{1+\numop{\delta_{\vec{k}}}}
\Big),
\label{eq:ssSublM}
\end{align}
from which the zero-temperature magnetization $m_A(T\!=\!0)$ and its temperature correction can be calculated; expressions for the $B$ sublattice follow analogously.



\begin{thebibliography}{99}
\bibitem{foot1}
The quoted interval $0.79<x<0.82$ \cite{yang21} deviates slightly from the value $x=0.765$ \cite{corboz13} shown in Fig.~\ref{fig:scboschem} because Refs.~\cite{corboz13} and \cite{yang21} employ different numerical methods.
\end{thebibliography}
\end{document}